\documentstyle[emulateapj,psfig,natbib,amsfonts,amsmath,graphicx]{article}
\def\tvlm{TVLM\,513-46546}
\def\swift{{\it Swift}}

\def\ociw{1}
\def\prince{2}
\def\hubble{3}
\def\udel{4}
\def\noao{5}
\def\mcgill{6}
\def\steward{7}
\def\iac{8}
\def\ucf{9}
\def\ucb{10}
\def\rice{11}

\begin{document}

\title{Simultaneous Multi-Wavelength Observations of Magnetic Activity
in Ultracool Dwarfs. I. The Complex Behavior of the M8.5 Dwarf
TVLM\,513-46546}

\author{
E.~Berger\altaffilmark{\ociw,}\altaffilmark{\prince,}\altaffilmark{\hubble},
J.~E.~Gizis\altaffilmark{\udel},
M.~S.~Giampapa\altaffilmark{\noao},
R.~E.~Rutledge\altaffilmark{\mcgill},
J.~Liebert\altaffilmark{\steward},
E.~Mart{\'{\i}}n\altaffilmark{\iac,}\altaffilmark{\ucf},
G.~Basri\altaffilmark{\ucb},
T.~A.~Fleming\altaffilmark{\steward},
C.~M.~Johns-Krull\altaffilmark{\rice},
N.~Phan-Bao\altaffilmark{\ucf},
W.~H.~Sherry\altaffilmark{\noao}
}

\altaffiltext{\ociw}{Observatories of the Carnegie Institution
of Washington, 813 Santa Barbara Street, Pasadena, CA 91101}
 
\altaffiltext{\prince}{Princeton University Observatory,
Peyton Hall, Ivy Lane, Princeton, NJ 08544}
 
\altaffiltext{\hubble}{Hubble Fellow}

\altaffiltext{\udel}{Department of Physics and Astronomy,
University of Delaware, Newark, DE 19716}

\altaffiltext{\noao}{National Solar Observatory, National Optical
Astronomy Observatories, Tucson, AZ 85726}

\altaffiltext{\mcgill}{Department of Physics, McGill University,
Rutherford Physics Building, 3600 University Street, Montreal,
QC H3A 2T8, Canada}

\altaffiltext{\steward}{Department of Astronomy and Steward
Observatory, University of Arizona, 933 North Cherry Avenue,
Tucson, AZ 85721}

\altaffiltext{\iac}{Instituto de Astrof{\'{\i}}sica de Canarias,
C/ V{\'{\i}}a L\'actea s/n, E-38200 La Laguna, Tenerife, Spain}

\altaffiltext{\ucf}{University of Central Florida, Department of
Physics, PO Box 162385, Orlando, FL 32816}

\altaffiltext{\ucb}{Astronomy Department, University of
California, Berkeley, CA 94720}

\altaffiltext{\rice}{Department of Physics and Astronomy, Rice
University, 6100 Main Street, MS-61 Houston, TX 77005}

\begin{abstract} 
We present the first simultaneous radio, X-ray, ultraviolet, and
optical spectroscopic observations of the M8.5 dwarf \tvlm, with a
duration of 9 hours.  These observations are part of a program to
study the origin of magnetic activity in ultracool dwarfs, and its
impact on chromospheric and coronal emission.  Here we detect steady
quiescent radio emission superposed with multiple short-duration,
highly polarized flares; there is no evidence for periodic bursts
previously reported for this object, indicating their transient
nature.  We also detect soft X-ray emission, with $L_X/L_{\rm
bol}\approx 10^{-4.9}$, the faintest to date for any object later than
M5, and a possible weak X-ray flare.  \tvlm\ continues the trend of
severe violation of the radio/X-ray correlation in ultracool dwarfs,
by nearly 4 orders of magnitude.  From the optical spectroscopy we
find that the Balmer line luminosity exceeds the X-ray luminosity by a
factor of a few, suggesting that, unlike in early M dwarfs,
chromospheric heating may not be due to coronal X-ray emission.  More
importantly, we detect sinusoidal H$\alpha$ and H$\beta$ equivalent
width light curves with a period of 2 hr, matching the rotation period
of \tvlm.  This is the first known example of such Balmer line
behavior, which points to a co-rotating chromospheric hot spot or an
extended magnetic structure, with a covering fraction of about $50\%$.
This feature may be transitory based on the apparent decline in light
curve peak during the four observed maxima.  From the radio data we
infer a large scale and steady magnetic field of $\sim 10^2$ G, in
good agreement with the value required for confinement of the X-ray
emitting plasma.  A large scale field is also required by the
sinusoidal Balmer line emission.  The radio flares, on the other hand,
are produced in a component of the field with a strength of $\sim 3$
kG and a likely multi-polar configuration.  The overall lack of
correlation between the various activity indicators suggests that the
short duration radio flares do not have a strong influence on the
chromosphere and corona, and that the chromospheric emission may not
be the result of coronal heating.
\end{abstract}
 

\section{Introduction}
\label{sec:intro}

In recent years it has become evident that low mass stars and brown
dwarfs (spectral classes late-M and L) are capable of producing
unanticipated levels of magnetic activity, manifested primarily in
their strong quiescent and flaring radio emission
\citep{bbb+01,ber02,brr+05,bp05,ber06,ohb+06,adh+07,pol+07,hbl+07,aob+07}.
Other traditional activity indicators such as chromospheric Balmer
line emission and coronal X-ray emission, however, appear to steeply
decline from peak levels in early and mid M dwarfs
\citep{pgr+81,vw87,gmr+00,whw+04}, leading to divergent trends of
magnetic activity at the bottom of the main sequence.  In both
H$\alpha$ and X-rays there is also a clear transition from persistent
emission to a small number of flaring objects with duty cycles of a
few percent \citep{rkg+99,gmr+00,rbm+00,lkc+03,whw+04}, as well as a
breakdown of the rotation-activity relation \citep{bm95,mb03} that is
clearly seen in early M dwarfs \citep{rgv85,fgs+93,mbs+02,pmm+03}.

The change and divergence in activity trends is most clearly evident
in the breakdown of the radio/X-ray correlation that holds for a large
number of early-type stars and solar flares \citep{gb93,gsb+93,bg94},
and is attributed to flare heating of coronal plasma to X-ray
temperatures \citep{neu68,gbs+96}.  While objects in the range M0--M6
obey this correlation, several objects later than M7 exhibit radio
emission that is several orders of magnitude brighter than expected
\citep{bbb+01,ber02,brr+05,ber06}.  Similarly, in early M dwarfs there
is an overall energy balance between X-ray and chromospheric emission,
which has led to the idea of chromospheric heating by coronal X-rays
(e.g., \citealt{cra82,hfs+95}).  It is not known whether this
mechanism holds in ultracool dwarfs, primarily because of the decline
in persistent activity.

Theoretical work on magnetic dynamos in ultracool dwarfs also remains
inconclusive.  Studies of the $\alpha^2$ dynamo in fully convective
stars suggest that a stratified and rotating turbulent medium can lead
to the build-up of a non-axisymmetric and multi-polar field (e.g.,
\citealt{ck06,dsb06}), but these models make several simplifications
for computational purposes.  It has also been argued that decreasing
electrical conductivity will impede the dissipation of any magnetic
fields in the cool and increasingly neutral atmospheres of ultracool
dwarfs \citep{mbs+02}.  The existing radio detections suggest that
field dissipation may not be a problem, but the overall field
configuration and the effect of neutral atmospheres in the presence of
magnetic dissipation remain largely unexplored.

As a result of the various conflicting trends and the transition from
persistent to flaring emission, progress in our understanding of
magnetic activity and dynamos in ultracool dwarfs requires {\it
simultaneous} observations of the various activity bands.  We have
already undertaken such observations for the L3.5 brown dwarf 2MASS
J00361617+1821104 in late 2002 and discovered periodic radio emission
($P=184$ min), with no corresponding X-ray or H$\alpha$ emission
\citep{brr+05}.  These observations indicated a magnetic field of
$\sim 200$ G covering a substantial fraction of the stellar surface,
as well as the first direct confirmation that the radio/X-ray
correlation is indeed violated by orders of magnitude.  Follow-up
observations showed that the field is stable on a $\gtrsim 3$ yr
timescale, much longer than the convective turnover time, pointing to
a stable dynamo process.  A recent simultaneous observation of the L
dwarf binary Kelu-1 resulted in an X-ray detection without
corresponding radio emission \citep{aob+07}, although the radio limits
still allow for a violation of the radio/X-ray correlation by up to
$\sim 2\times 10^3$.

Here we exploit the powerful approach of simultaneous observations to
investigate the magnetic activity in the M8.5 dwarf \tvlm, an object
previously detected in the radio
\citep{ber02,ber06,ohb+06,had+06,hbl+07} and in H$\alpha$
\citep{mrm94,rkl+02,mb03}.  These observations are the first in a
series that targets several objects in the sparsely-studied and
critical spectral type range M7 to L3.  In the case of \tvlm\ we
detect radio, X-ray, and Balmer line emission, but no UV emission.
The overall behavior is complex and largely uncorrelated between the
various emission bands.  The long time and wavelength baselines
provide unprecedented detail, including the first case to date of
sinusoidal Balmer line emission; the observed 2-hour period is in
excellent agreement with the rotation of \tvlm.  Using the various
activity indicators we infer the properties of the magnetic field,
corona, and chromosphere, and show that the underlying processes and
field configuration likely differ from those in early M dwarfs.

\section{Observations}
\label{sec:obs}

We targeted the M8.5 dwarf \tvlm\ due to its vicinity ($d=10.6$ pc;
\citealt{dhv+02}) and known radio and H$\alpha$ activity.  The
bolometric luminosity of \tvlm\ is $L_{\rm bol}\approx 10^{-3.59}$
L$_\odot$, and its rotation velocity is $v{\rm sin}i\approx 60$ km
s$^{-1}$ \citep{mb03}.  The non-detection of lithium, with a limit of
0.05 \AA, suggests that \tvlm\ is most likely a very low mass star
\citep{rkl+02}.  Adaptive optics imaging of \tvlm\ revealed no
companions with $\delta m\lesssim 3$ mag in the range
$0.1$-$15\arcsec$ \citep{csf+03}.

\tvlm\ was first detected in the radio during a 2 hr observation at
8.5 GHz in Sep.~2001, and exhibited both persistent ($F_\nu\approx
190$ $\mu$Jy) and flaring emission \citep{ber02}, the latter with a
peak brightness of 1 mJy, a duration of 15 min, and circular
polarization of $r_c\approx 66\%$.  Subsequent observations from 1.4
to 8.5 GHz in Jan.~2004 revealed a similar level of persistent
emission, $F_\nu(1.4)\approx 260$ $\mu$Jy, $F_\nu(4.9)\approx 280$
$\mu$Jy, and $F_\nu(8.5)\approx 230$ $\mu$Jy, with $r_c\lesssim 15\%$
\citep{ohb+06}.  Observations in Jan.~2005 revealed brighter emission,
$F_\nu(4.9)\approx 405$ $\mu$Jy, and $F_\nu(8.5)\approx 400$ $\mu$Jy,
as well as a claimed periodicity of about 2 hr \citep{had+06}.
Finally, observations in May 2006 uncovered a series of flares with
durations of a few minutes, $\sim 100\%$ circular polarization, and a
periodicity of 1.96 hr \citep{hbl+07}.

Previous detections of H$\alpha$ emission reveal long term
variability, with equivalent widths (EW) ranging from 1.7 to 3.5 \AA,
or $L_{\rm H\alpha}/L_{\rm bol}\approx 10^{-5}$
\citep{mrm94,rkl+02,mb03}.

Our simultaneous observations were obtained on 2007 April 20 UT for a
total of $8.8$ hr in the radio (04:00-12:48 UT), $8.9$ hr in the
X-rays (03:47-12:41 UT), and $7$ hr in the optical (07:13-14:13 UT).
{\it Swift} UV/optical telescope (UVOT) observations took place
intermittently between 03:58 UT and 12:10 UT, with a total on-source
exposure time of 8036 s.

\subsection{Radio}
\label{sec:rad}

Very Large Array\footnotemark\footnotetext{The VLA is operated by the
National Radio Astronomy Observatory, a facility of the National
Science Foundation operated under cooperative agreement by Associated
Universities, Inc.} observations were conducted at a frequency of 8.46
GHz in the standard continuum mode with $2\times 50$ MHz contiguous
bands.  Scans of 295 s on source were interleaved with 50 s scans on
the phase calibrator J1513+236. The flux density scale was determined
using the extragalactic source 3C\,48 (J0137+331).

The data were reduced and analyzed using the Astronomical Image
Processing System (AIPS). The visibility data were inspected for
quality, and noisy points were removed.  To search for source
variability, we constructed light curves using the following
method. We removed all the bright field sources using the AIPS/IMAGR
routine to CLEAN the region around each source, and the AIPS/UVSUB
routine to subtract the resulting source models from the visibility
data.  We then plotted the real part of the complex visibilities at
the position of \tvlm\ as a function of time using the AIPS/DFTPL
routine.  The subtraction of field sources is required since their
sidelobes and the change in the shape of the synthesized beam during
the observation result in flux variations over the map that may
contaminate real variability or generate false variability.  The
resulting light curves are shown in Figures~\ref{fig:all} and
\ref{fig:flares}.

\subsection{X-Rays}
\label{sec:xrays}

The observations were made with the Chandra/ACIS-S3
(backside-illuminated chip), with \tvlm\ offset from the on-axis focal
point by $15''$.  A total of 29.76 ks were obtained.  Data were
analyzed using CIAO version 3.3, and counts were extracted in a $1''$
radius circle centered on the source position.  We find a total of 8
counts in the $0.2-2$ keV range, and 2 additional counts with
$kT\approx 10$ keV.  Background counts were extracted from annuli
centered on the source position, excluding other point sources
detected in the observation.  We find that 2 background counts are
expected within the source extraction aperture, likely corresponding
to the two photons with $kT\approx 10$ keV.

The source counts exhibit a narrow energy range with $\langle
kT\rangle=930\pm 250$ eV, corresponding to a typical plasma
temperature of $1.1\times 10^7$ K.  Using this temperature with a
Raymond-Smith plasma model we find an energy conversion factor of
$1\,{\rm count}=3.4\times 10^{-12}$ erg cm$^{-2}$ s$^{-1}$ ($0.2-2$
keV).  Thus, the observed count rate of $2.69\times 10^{-4}$ s$^{-1}$
translates to a flux of $9.3\times 10^{-16}$ erg cm$^{-2}$ s$^{-1}$
($0.2-2$ keV), or a flux density of $1.4$ nJy at $kT=1$ keV.  At the
distance of \tvlm\ the corresponding luminosity is $L_X\approx
1.2\times 10^{25}$ erg s$^{-1}$, or a ratio of $L_X/L_{\rm bol}\approx
10^{-4.9}$.  This detection is at the same level as the quiescent
emission from the M8 dwarf VB\,10 \citep{fgg03}, the faintest X-ray
emitting late-M dwarf to date.

We next find that of the 8 detected photons 4 arrive as pairs with
separations of 217 and 31 s (Figure~\ref{fig:all}).  The chance
probabilities of such short time separations in a 29.76 ks observation
are $1.7\times 10^{-3}$ and $3.4\times 10^{-5}$, respectively.  It is
thus possible that the second pair constitutes a flare.  If true, the
flare luminosity is $3.1\times 10^{24}$ erg cm$^{-2}$ s$^{-1}$, or
$L_X/L_{\rm bol}\approx 10^{-5.5}$, the lowest luminosity flare
detected from any late-M dwarf to date.  The quiescent component would
be correspondingly lower, $L_X/L_{\rm bol}\approx 10^{-5.0}$.  As we
show below, the putative X-ray flare may coincide with the peak of the
broadest radio flare.

\subsection{Optical Spectroscopy} 
\label{sec:optical}

We used\footnotemark\footnotetext{Observations were obtained as part
of program GN-2007A-Q-60.} the Gemini Multi-Object Spectrograph (GMOS;
\citealt{hja+04}) mounted on the Gemini-North 8-m telescope with the
B600 grating set at a central wavelength of 5250 \AA, and with a $1''$
slit.  A series of eighty 300-s exposures were obtained with a readout
time of $18$ s providing $94\%$ efficiency.  The individual exposures
were reduced using the {\tt gemini} package in IRAF (for bias
subtraction and flat-fielding), and rectification and sky subtraction
were performed using the method and software described in
\citet{kel03}.  Wavelength calibration was performed using CuAr arc
lamps and air-to-vacuum corrections were applied.  The spectrum covers
$3840-6680$ \AA\ at a resolution of about 5 \AA.

\begin{figure}
\centerline{\psfig{file=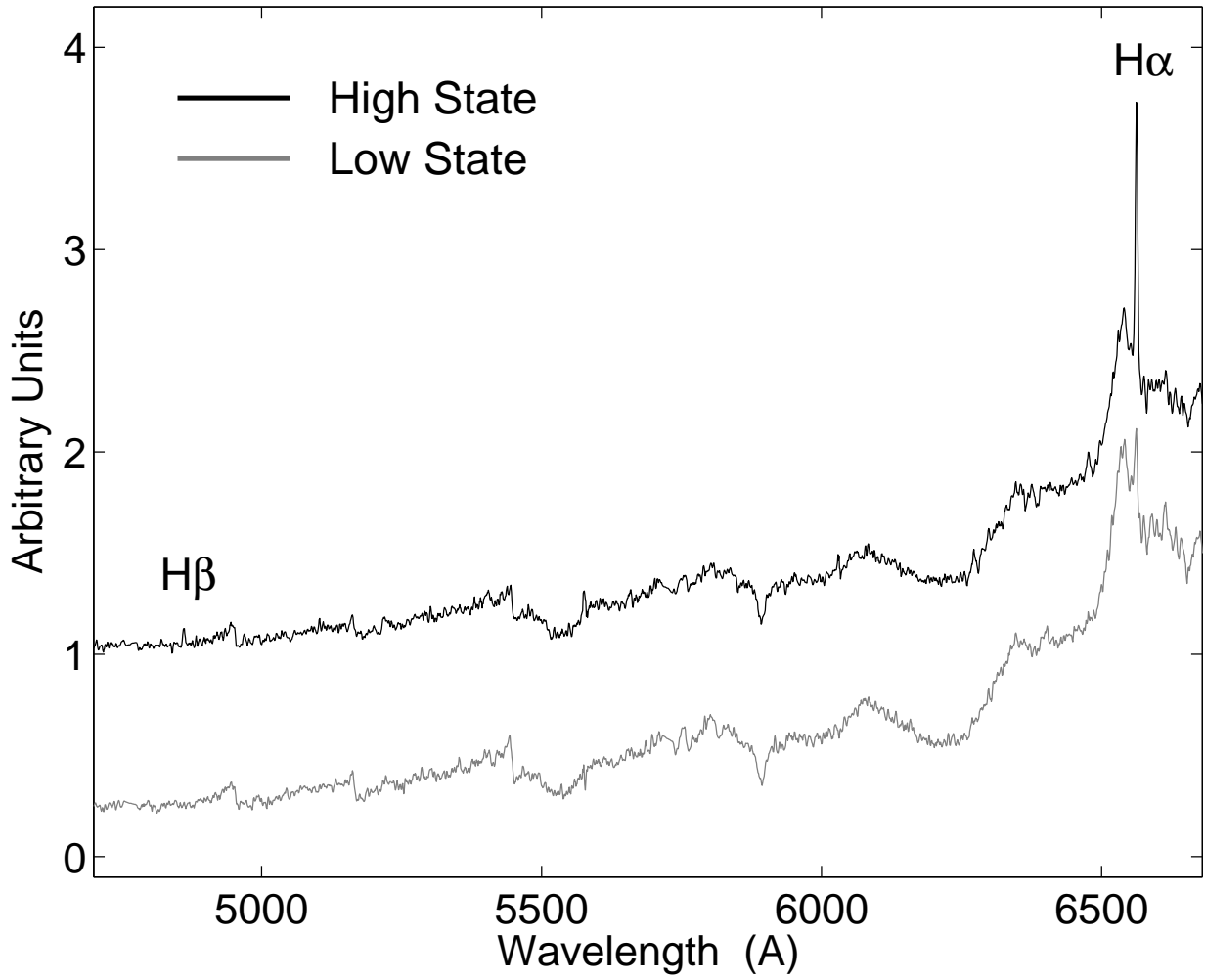,width=6.5in}}
\caption{Sample optical spectra of \tvlm\ in the high and low Balmer 
emission line states.  The high state spectrum has been offset upward
for clarity.
\label{fig:optical}}
\end{figure}

To measure the equivalent widths of the H$\alpha$ and H$\beta$
emission lines we use continuum regions centered on 6551 and 6572 \AA,
and on 4854 and 4870 \AA, respectively.  Sample spectra in the low and
high Balmer emission state are shown in Figure~\ref{fig:optical}.  The
H$\alpha$ light curve exhibits a clear sinusoidal behavior
(Figure~\ref{fig:all}).

\subsection{Ultraviolet}

The data were obtained with the \swift/UVOT in the UVW1 filter
($\lambda_{\rm eff}\approx 2510$ \AA), as a series of 6 images with
exposure times ranging from 560 to 1630 s (Figure~\ref{fig:all}).  No
source is detected at the position of \tvlm\ in any of the individual
exposures, or in the combined image with a total exposure time of 8036
s.  We performed photometry on the combined exposure using a circular
aperture matched to the PSF of the UVW1 filter ($2.2''$), and found a
$3\sigma$ limit of $F_\lambda({\rm UVW1})<2.4\times 10^{-18}$ erg
cm$^{-2}$ s$^{-1}$ \AA$^{-1}$, or a Vega magnitude of $m({\rm
UVW1})>23.0$ mag.  This limit corresponds to a ratio of UV to
bolometric luminosity of $\lambda L_\lambda/L_{\rm bol}<10^{-3.2}$.

\section{Multi-Wavelength Emission Properties} 
\label{sec:prop}

We observed \tvlm\ across a wide wavelength range that traces activity
in various layers of the outer atmosphere.  The radio emission traces
particle acceleration by magnetic processes, and corresponds to
gyrosynchrotron radiation or coherent radiation (electron cyclotron
maser or plasma emission).  The Balmer emission lines are thought to
be collisionally excited in the chromosphere, and the X-ray thermal
emission arises in the corona.

\subsection{Quiescent Emission}
\label{sec:quiet}

Using the observed X-ray flux and temperature in the context of
bremsstrahlung radiation we can estimate the coronal gas density and
pressure.  The emissivity is given by:
\begin{equation}
\eta_\nu\approx 7.6\times 10^{-38}\,{\rm ln}(T/\nu)\,n_e^2\,T^{-1/2}
\,\,\, {\rm erg\,s^{-1}\,cm^{-3}\,Hz^{-1}},
\end{equation}
where for \tvlm, $T=1.1\times 10^7$ and $\nu=2.2\times 10^{17}$ Hz.
Assuming a uniform coronal structure extending $\approx (0.1-1)
R_*\approx (0.7-7)\times 10^9$ cm above the photosphere, we find
$F_{\nu,X}\approx (0.3-6)\times 10^{-52}\,n_e^2$ erg s$^{-1}$
cm$^{-2}$ Hz$^{-1}$.  From the observed flux density of 1.4 nJy we
thus infer an electron density, $n_e\approx (0.5-2)\times 10^{10}$
cm$^{-3}$.  We note that a coronal structure with a low volume filling
factor, $f$, would have a correspondingly higher density, $n_e\propto
f^{-1/2}$.  Given this overall uncertainty, the inferred density is
similar to what has been found for the M4.5 flare star AD Leo (flares:
$n_e\sim 10^{10}-10^{11}$ cm$^{-3}$, quiescent: $n_e<3\times 10^{10}$
cm$^{-3}$; \citealt{vrm+03}) and is somewhat lower than in the M3.5
flare star EV Lac ($n_e\sim 10^{13}$ cm$^{-3}$; \citealt{oha+06}).

From the temperature and inferred density we find a coronal gas
pressure of $P\approx 10-40$ dyne cm$^{-2}$.  This does not take into
account turbulent pressure, which at coronal temperatures may in fact
be dominant (e.g., \citealt{oha+06}.  If the corona is confined by
magnetic fields, the required average field strength is thus
$B>\sqrt{8\pi P}\gtrsim 15-30$ G.

An alternative estimate of the coronal physical parameters can be
provided in the context of the loop model calculated by
\citet{rtv78}.  In this model the loop temperature, density, and 
pressure are related by $T\approx 1.4\times 10^3\,(P\ell)^{1/3}$ K,
where it is assumed that the loop apex temperature is similar to the
mean coronal temperature.  The loop length, $\ell$, is determined by
the observed X-ray luminosity, $\ell\approx 2.2\times 10^{14}\,f\,
T^{7/2}\,L_X^{-1}$ cm, where $f$ is the filling factor of the loops.
The simple relation between $T$, $P$, and $\ell$ holds as long as
$P\sim {\rm const}$ along the loop.  This requires the loop length to
be less than the pressure scale height, $H\approx 7.5\times 10^2\,T$
cm (for $R=0.1$ R$_\odot$ and $M=0.08$ M$_\odot$).  From the observed
parameters we find $H\approx 8\times 10^9$ cm and $\ell\approx 8\times
10^{13}\,f$ cm.  The condition $\ell<H$ thus requires a very small
filling factor, $f\lesssim 10^{-4}$.  The pressure inferred from using
$\ell\sim H$ is $P\approx 60$ dyne cm$^{-2}$, and the magnetic field
required for confinement is thus $B\gtrsim 40$ G, in good agreement
with the value inferred above by assuming a uniform coronal structure.

We now turn to the radio emission.  The bremsstrahlung contribution in
the radio band is $F_\nu\approx 0.03$ $\mu$Jy, orders of magnitude
lower than even the baseline quiescent emission, which from several
flare-free regions of the radio light curve is found to be $F_\nu
(8.5)=208\pm 18$ $\mu$Jy.  The $3\sigma$ limit on the fraction of
circular polarization of the quiescent component is $r_c<25\%$.  Both
the flux and degree of circular polarization are similar to those
measured in previous observations (\S\ref{sec:obs}), indicating that
the quiescent component is stable on a multi-year timescale.

Based on the brightness of the radio emission compared to the
predicted thermal emission, and its long term stability, we conclude
that it is most likely due to gyrosynchrotron radiation.  We follow
the typical assumption that the mildly relativistic electrons, which
produce the radio emission, follow a power law distribution,
$N(\gamma)\propto\gamma^{-p}$ for $\gamma>\gamma_m$, with $p\sim 3$
typical for M and L dwarfs \citep{gsb+03,brr+05,ohb+06}.  The
gyrosynchrotron emission spectrum is determined by the size of the
emission region ($R$), the density of radiating electrons ($n_e$), and
the magnetic field strength ($B$) according to
\citep{dm82}:
\begin{equation}
r_c=0.3\times 10^{1.93{\rm cos}\theta-1.16{\rm cos}^2\theta} 
(3\times 10^3/B)^{-0.21-0.37{\rm sin}\theta},
\end{equation}
\begin{equation}
\nu_m=1.8\times 10^4\,({\rm sin}\theta)^{0.5}\,(n_eR)^{0.23}
\,B^{0.77}\,\,\,{\rm Hz},
\end{equation}
\begin{equation}
F_{\nu,m}=2.5\times 10^{-41}\,B^{2.48}\,R^3\,n_e\,({\rm 
sin}\theta)^{-1.52}\,\,\,{\rm \mu Jy},
\end{equation}
where $\theta$ is the angle between the magnetic field and the line of
sight.

From previous observations of \tvlm\ it appears that the peak of the
quiescent emission spectrum is $\nu_m\approx 5$ GHz \citep{ohb+06}.
Using the limit $r_c<25\%$, we infer a magnetic field strength, $B
\lesssim 10-740$ G, for $\theta=20-80^\circ$.  With this range we find
$R\sim (0.8-8)\times 10^10$ cm, and $n_e\sim 100-1.3\times 10^{10}$
cm$^{-3}$; the latter range is for $\theta=80^\circ$ to $20^\circ$.  A
comparison to the coronal density estimated from X-rays indicates
$\theta\sim 20-30^\circ$, and hence $R\sim {\rm few}\times 10^{10}$ cm
and $B\lesssim {\rm few}\times 10^2$ G.  The inferred field strength
is consistent with the value required for confinement of the X-ray
emitting coronal plasma as derived above.

Finally, we turn to the observed H$\alpha$ emission.  The light curve
exhibits clear sinusoidal behavior with a range of equivalent widths
of $1.5-5.5$ \AA\ (Figure~\ref{fig:all}).  The sinusoidal behavior
indicates that the line is likely modulated by the rotation of \tvlm\
rather than flares, and the emission is thus persistent in origin, at
least on the timescale of our observation.  \citet{whw04} determined a
multiplicative factor, $\chi\equiv f_{\lambda 6560}/f_{\rm bol}$, to
convert H$\alpha$ EW to $L_{\rm H\alpha}/L_{\rm bol}$.  For \tvlm, the
observed $I-K=4.3$ mag indicates ${\rm log}\chi\approx -5.3$, which
matches the average value for spectral type M8.5 \citep{whw04}.  Thus,
for the full range of EWs we find ${\rm log}(L_{\rm H\alpha}/L_{\rm
bol})\approx -4.6$ to $-5.1$.  This covers the typical range of
H$\alpha$ emission observed from M8.5 dwarfs \citep{whw+04}, as well
as the range of EWs from past observations of this source
(\S\ref{sec:obs}).  We return to the implications of the sinusoidal
variations in \S\ref{sec:halpha}.

Since the X-ray and H$\alpha$ emission appear to be persistent, we can
gain insight into the physical processes in the outer atmosphere by
examining the energy scale in each band.  In early dMe stars there is
evidence that the quiescent chromosphere is heated by downward
directed coronal X-ray emission \citep{cra82}.  Nearly half of the
absorbed energy will be radiatively lost in the Balmer lines,
primarily H$\alpha$ \citep{cra82}.  Here we find that the ratio
$L_{\rm H\alpha}/L_X\approx 2$ is in rough agreement with this overall
model, taking into account the overall uncertainties in the X-ray and
H$\alpha$ luminosities.  However, as noted by \citet{cra82}, any
significant asymmetry in the coronal structure (e.g., extended loops)
will substantially reduce the X-ray flux directed at the chromosphere.
Since we find that the H$\alpha$ line by itself already accounts for
the maximal coronal heating rate, this model of chromospheric heating
requires a uniform coronal structure.

Alternatively, the marginal excess chromospheric luminosity may
indicate that the structure of the outer atmosphere in this late-M
dwarf is markedly different than in early dMe stars.  Potentially, the
H$\alpha$ emission is dominated by a non-chromospheric component.  As
we discuss below, in the context of the periodic emission, it is
possible that the H$\alpha$ emission region is a more extended
``bubble'' that co-rotates with the star.

\subsection{Radio Flares}
\label{sec:flare}

\begin{figure}
\centerline{\psfig{file=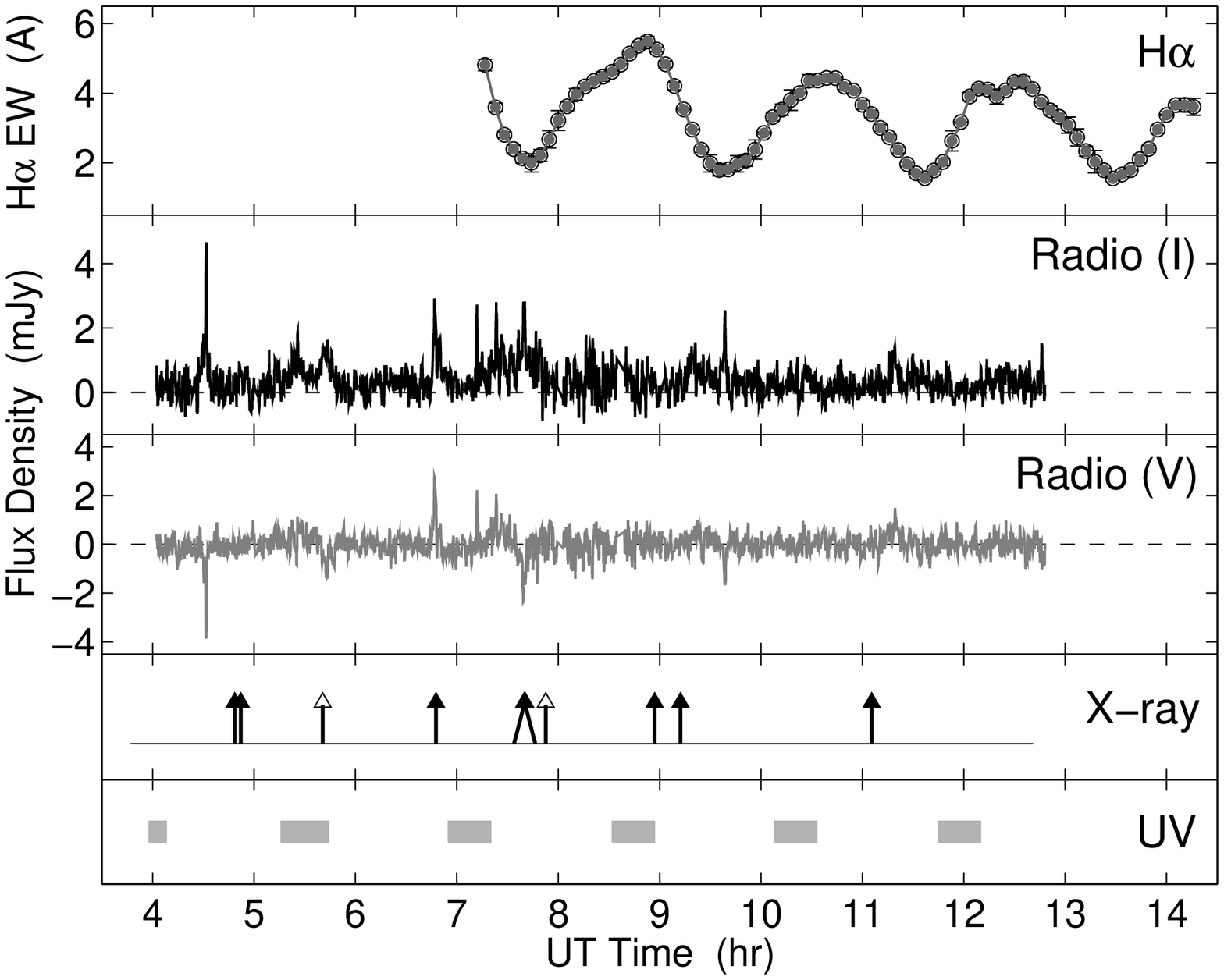,width=6.5in}}
\caption{Radio, H$\alpha$, UV, and X-ray of \tvlm.  The arrival
times of the X-ray photons are shown with arrows.  Empty arrows
correspond to likely background events with $kT\gtrsim 10$ keV; two
are expected from the background count rate.  Times of UV coverage are
marked by gray squares; no UV emission is detected.  The H$\alpha$
emission is clearly sinusoidal and periodic, with $P\approx 2$ hr
matching the rotation period of \tvlm, with an implied ${\rm
sin}i\approx 1$.  There is no clear correspondence between the various
emission bands, with the possible exception of an X-ray photon pair
that coincides with the broadest radio flare (at about 07:40 UT).
\label{fig:all}}
\end{figure}

\begin{figure}
\centerline{\psfig{file=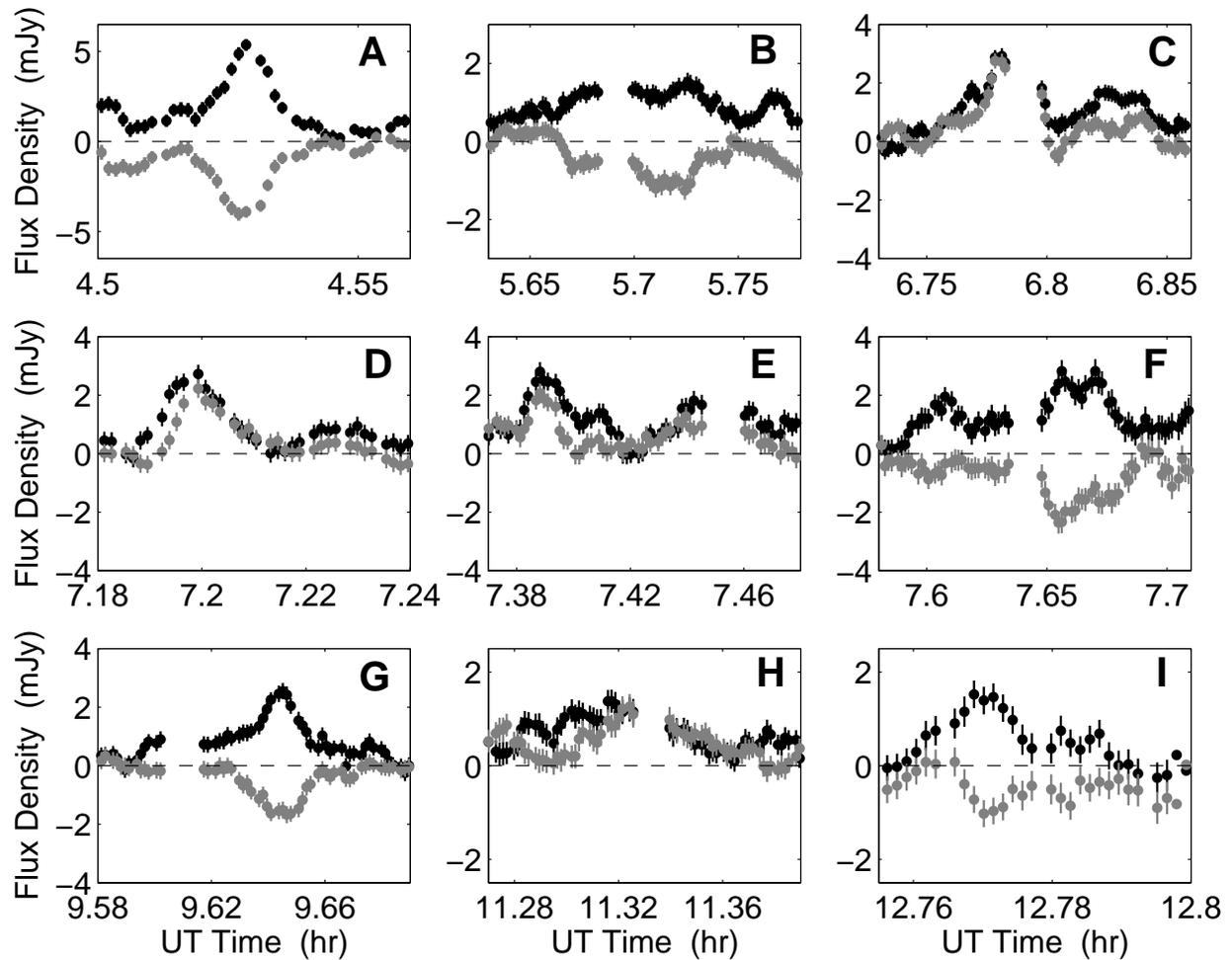,width=6.5in}}
\caption{Zoom-in on individual radio flares.  Total intensity (black) 
and circularly polarized flux (gray) are shown.  The flares exhibit
diverse behavior in terms of duration, amplitude, and fraction and
sense of circular polarization.
\label{fig:flares}}
\end{figure}

In addition to the persistent emission in the radio band we detect
several distinct short duration ($\delta t\sim 2-15$ min) flares, and
a single broad brightening with a duration of about 1 hr
(Figures~\ref{fig:all} and \ref{fig:flares}).  These flares range in
peak flux density from $2$ to $5.5$ mJy.  Our observations allow the
detection of flares to a $5\sigma$ sensitivity of about $3.5$ mJy in a
single 5 s integration.  In Figure~\ref{fig:flares} we provide a
zoom-in on the most distinct flares, including both the total
intensity light curve and the circular polarization light curve.
About half of the flares exhibit left circular polarization, while the
other half exhibit right circular polarization, with overall fractions
of $\sim 50-100\%$.  There is no apparent regularity in the duration,
brightness, arrival time, or circular polarization of the flares.

The apparently random sense of circular polarization (right-
vs.~left-handed) and flare arrival times suggest that the flares are
produced in unrelated regions.  This is contrary to the model proposed
by \citet{hbl+07} to explain their detection of periodic flares, with
$P=1.96$ hr well-matched to the rotation of \tvlm.  These authors
point to emission from a distinct region that is stable over at least
several rotation periods.  The absence of a similar behavior in our
data suggests that this stability is limited to $\lesssim 1$ yr.

The variation in the sense of circular polarization from one flare to
the next is also different from the uniform sense of polarization
observed in radio flares from UV Cet (M5.5) and YZ CMi (M4.5).  Flares
in these two stars show consistent right-handed and left-handed
polarization, respectively, which has been interpreted as a signature
particle acceleration in a toroidal dipole field \citep{kbc+02}.  The
behavior of radio flares from \tvlm\ in the current observation thus
points to a multi-polar field configuration.

While the quiescent radio emission is due to gyrosynchrotron
radiation, the short durations and large fraction of circular
polarization of the flares point to emission by coherent processes.
The electron cyclotron maser (ECM) process leads to emission at the
fundamental cyclotron frequency, $\nu_c=2.8\times 10^6\,B$ Hz, while
plasma radiation is dominated by the fundamental plasma frequency,
$\nu_p=9\times 10^3\,n_e^{1/2}$ Hz.  In the ECM case the magnetic
field strength inferred from the detection of flares at 8.46 GHz is
$B\approx 3$ kG, while in the case of plasma radiation we infer an
electron density of $n_e\approx 9\times 10^{11}$ cm$^{-3}$.  The
latter should be used as an upper limit in the ECM model.

In the standard picture of the electron cyclotron maser (e.g.,
\citealt{md82}), electrons with a large pitch angle are reflected at
the legs of the magnetic loop, while those with smaller angles
precipitate into and heat the chromosphere.  This process repeats,
with the rise time of the emitted flare presumably reflecting the
particle acceleration timescale.  The flare decay reflects the time to
precipitate out of the magnetic trap, and is thus related to the loss
rate, $\tau\sim [(\Omega_L/4\pi)v/\ell]^{-1}$, where $\Omega_L$ is the
solid angle of the loss cones, $v$ is the electron velocity, and
$\ell$ is the transit length \citep{md82}.  From our observations at
8.5 GHz we infer $v\sim c$, and hence $\tau\sim 30$ s leads to,
$\ell\sim 10^{12}\,(\Omega_L/4\pi)$ cm.  The actual size of the loop
region is likely to be much smaller since the accelerated electrons
traverse the field lines multiple times before the maser process
ceases.  Thus, whereas in the case of the periodic outbursts a stable,
dipolar field configuration was inferred \citep{hbl+07}, here it is
likely that the emission is arising from a tangled, multi-polar field.

Finally, the ECM processes is supposed to provide coronal and
chromospheric heating as electrons precipitate out of the trap, and
thus lead to increased H$\alpha$ and X-ray emission.  From
Figure~\ref{fig:all} it is clear that this is not the case for \tvlm.
None of the flares radio are accompanied by an increase in either the
H$\alpha$ or X-ray emission.  This may provide further support to the
proposed tangled field configuration, since it will lead to
chromospheric and coronal effects on a small scale that may go
undetected when averaged over the stellar disk.

\subsection{Sinusoidal H$\alpha$ Emission}
\label{sec:halpha}

As noted above the H$\alpha$ and H$\beta$ emission lines exhibit
sinusoidal variations with a periodicity of about 2 hours
(Figure~\ref{fig:all}).  This indicates that the likely origin of the
Balmer line modulation is the rotation of \tvlm.  Indeed with
$R_*\approx 7\times 10^9$ cm, the 2-hour period requires a rotation
velocity of $v\approx 60$ km s$^{-1}$, in perfect agreement with the
observed $v{\rm sin}i\approx 60$ km s$^{-1}$.  If the Balmer lines are
modulated by rotation, this indicates an inclination of the rotation
axis relative to the line of sight of $i\approx 90^\circ$.

The likely connection to stellar rotation points to a large scale
($f\sim 50\%$) hot spot located on one hemisphere of \tvlm, or an
emission ``bubble'' that may extend to several times the stellar
radius, and is occulted by the star for about half a rotation period.
As noted above, the latter scenario may be relevant since the observed
H$\alpha$ luminosity appears to exceed the X-ray luminosity by a
factor of a few, raising the possibility of an H$\alpha$ emission
region distinct from the chromosphere.  The fact that the Balmer lines
do not completely disappear during the light curve minima suggests
that in both scenarios the solid angle subtended by the emission
region is larger than the stellar disk, or alternatively, that the
minima represent the baseline chromoshperic emission, while the
rotation of an extended bubble or a hot spot into our line of sight
produces the maxima.

It also appears from the H$\alpha$ light curve that the peak
equivalent width decreases with time (Figure~\ref{fig:all}).  The
first peak has ${\rm EW}\approx 5.5$ \AA\, followed by about 4.5 \AA\
for the two subsequent peaks, and about 3.8 \AA\ for the fourth peak.
While our observations start during the decline of a preceding cycle,
an extrapolation back to the time of the expected peak indicates an
equivalent width of at least 6 \AA.  The equivalent widths at the
minima, however, appear to be more stable, with perhaps a slight
decrease from 2 to 1.5 \AA\ during our observation.  These trends
suggest that the hot spot or extended bubble may be transitory,
possibly excited by an energetic event prior to the beginning of our
observations.  Extrapolating the observed trend, we find that the peak
H$\alpha$ equivalent width will match the baseline level of 1.5 \AA\
about 7 hrs after the end of our observation.  Future observations of
\tvlm\ are crucial for assessing whether this is in fact a transient 
feature.

We further find no clear correspondence between the peaks of H$\alpha$
emission and the radio flares.  In fact, the opposite may be the case.
The three H$\alpha$ minima that have simultaneous radio coverage
appear to roughly coincide with radio flares.  If we extrapolate the
H$\alpha$ light curve back in time, we find that a previous minimum
also appears to coincide with a radio flare (at about 05:40 UT).
However, other radio flares, including the brightest one detected, do
not coincide with observed or extrapolated H$\alpha$ minima,
suggesting that this anti-correlation between H$\alpha$ and radio
emission may be a result of small number statistics.

The detection of a 2 hour period in two different emission bands and
at two different times with no clear correspondence indicates that the
radio flares are not sufficiently energetic, or impact a large enough
scale to influence the chromosphere.  This behavior also suggests that
the magnetic field configuration of \tvlm\ may be composed of a
large-scale dipolar field, as well as a more compact and tangled
component, which give rise to periodic signals at different times and
wavebands.

\section{Lack of Radio/X-ray Correlation}
\label{sec:trend}

\begin{figure}
\centerline{\psfig{file=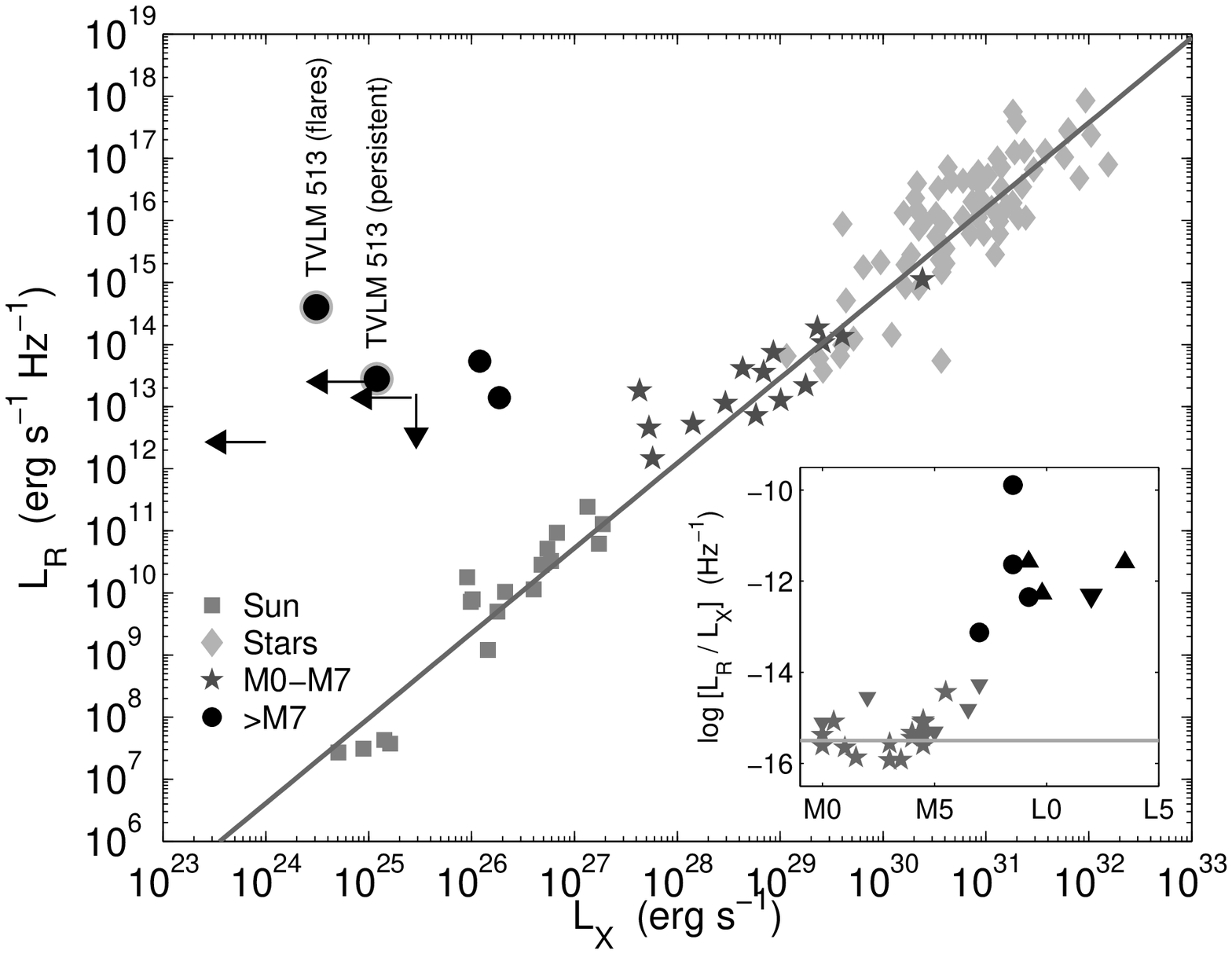,width=6.5in}}
\caption{Radio vs. X-ray luminosity for stars exhibiting coronal 
activity.  Data for late-M and L dwarfs are from \citet{rbm+00},
\citet{bbb+01}, \citet{ber02}, \citet{brr+05}, \citet{bp05}, 
\citet{ber06}, and \citet{aob+07}, while other data are taken from 
\citet{gud02} and references therein.  Data for the Sun include 
impulsive and gradual flares, as well as microflares.  The strong
correlation between $L_R$ and $L_X$ is evident, but it begins to break
down around spectral type M7 (see inset).
\label{fig:gb}}
\end{figure}

\tvlm\ is only the third ultracool dwarf to be observed 
simultaneously in the radio and X-rays, allowing a continued
investigation of the radio/X-ray correlation in these objects.
Coronally active stars up to spectral type M7, including the Sun,
exhibit a tight correlation between their radio and X-ray emission
\citep{gb93,bg94}.  The persistent emission follows a linear trend,
$L_R\approx 3\times 10^{-16}\,L_X$ Hz$^{-1}$, which extends over 6
orders of magnitude in $L_R$ \citep{gb93}, while for flares the
relation is $L_R\approx 5.4\times 10^{-27}\,L_X^{1.37}$ Hz$^{-1}$ over
8 orders of magnitude in $L_R$ \citep{bg94}; see Figure~\ref{fig:gb}.

From several previous (non-simultaneous) observations it has become
clear that the above correlations break down in at least some
ultracool dwarfs \citep{bbb+01,ber02,brr+05,ber06}, with a clear
transition occurring at spectral type M7 \citep{ber06}.  This
conclusion was supported by our simultaneous observations of the L3.5
dwarf 2MASS J00361617+1821104 \citep{brr+05}.  Here we find continued
evidence for this trend.  Using the persistent radio and X-ray
luminosities we find $L_R/L_X\approx 2.3\times 10^{-12}$ Hz$^{-1}$, a
factor of about 7800 times larger than expected.  This is similar to
the level of excess radio emission observed in the previous late-M and
L dwarfs.  If we take into account the flaring emission, with a
typical peak luminosity of $L_R\approx 4\times 10^{14}$ erg cm$^{-2}$
s$^{-1}$ Hz$^{-1}$, and assume that the single X-ray photon pair with
$L_X\approx 3\times 10^{24}$ erg cm$^{-2}$ s$^{-1}$ is indeed a flare
(\S\ref{sec:xrays}), we find that for flares $L_R/L_X\approx 1.3\times
10^{-10}$ Hz$^{-1}$, a factor of about $10^7$ larger than expected.

The radio/X-ray correlation has been interpreted in the context of the
Neupert effect \citep{neu68}. In this scenario, the radio emission is
produced when coronal magnetic loops reconnect and create a current
sheet along which ambient electrons are accelerated.  The accelerated
electrons in turn drive an outflow of hot plasma into the corona as
they interact with and evaporate the underlying chromospheric
material.  The interaction of the outflowing plasma with the electrons
produces X-ray emission via the bremsstrahlung process.  This
mechanism points to a causal connection between particle acceleration,
which is the source of radio emission, and plasma heating, which
results in X-ray emission. Thus, the X-ray thermal energy should
simply be related by a constant of proportionality to the integrated
radio flux.  The breakdown in the radio/X-ray correlation, and the
clear lack of correspondence between the radio flares observed on
\tvlm\ and its X-ray emission suggest that the heating of coronal
material is generally inefficient in objects later than M7.  This may
be due to the short lifetime and small size of the flare emitting
regions (i.e., in the case that they arise from highly tangled and
multi-polar fields), or to lower efficiency of coronal heating so that
the bulk of the coronal emission is at temperatures lower than $kT\sim
1$ keV.

\section{Discussion and Conclusions}
\label{sec:conc}

We presented simultaneous radio, X-ray, UV, and optical spectroscopic
observations of the M8.5 dwarf \tvlm\ that probe magnetic activity and
its influence on the stellar chromosphere and corona.  These
observations are the first in a series of several objects that span
the sparsely-studied spectral type range M7--L3, over which the
magnetic activity appears to exhibit a change in behavior compared to
early spectral types.  We find that \tvlm\ exhibits a wide range of
both quiescent and flaring activity that includes radio emission from
large and small scale regions, coronal soft X-ray emission, and
sinusoidal and periodic Balmer line emission from a chromospheric hot
spot or an extended structure with a covering fraction of about
$50\%$.

Quantitatively, we find that the quiescent radio emission is produced
on the scale of the entire stellar disk, with a magnetic field of
$\sim 10^2$ G.  From portions of the light curve that are free of
flares we see no evidence for variability of this persistent
component.  Moreover, the similarity in flux level to observations
carried over the past 6 years indicates that the magnetic field is
stable on timescales significantly longer than the convective turnover
time, $\tau_{\rm conv}\sim 10^2$ d \citep{cb00}.  This is similar to
the situation we observed in 2MASS J00361617+1821104 (L3.5), with a
periodic radio signal that was stable over at least 3 years
\citep{brr+05}.

We also find from a comparison of the radio and X-ray data that the
magnetic field axis is likely highly inclined relative to the rotation
axis of \tvlm, which is inferred from the 2-hour period of the
H$\alpha$ emission to be about $90^\circ$ (\S\ref{sec:halpha}).  This
is an interesting result in the context of magnetic dynamo models of
fully convective stars.  \citet{ck06} and \citet{dsb06} found that the
$\alpha^2$ dynamo, which relies on a stratified and rotating turbulent
medium, leads to a non-axisymmetric field with an overall configuration
that lies in the equatorial plane.  This seems to be supported by our
observations.

The X-ray emission requires a corona with $T\approx 10^7$ K and a
density of $n_e\sim 10^{10}$ cm$^{-3}$, similar to those of early M
dwarfs.  The inferred coronal gas pressure requires a magnetic field
strength of at least $\sim 10-40$ G for confinement, in good agreement
with the radio-derived field strength.  The energy input from the
X-ray emitting corona is similar to, or somewhat smaller than, the
radiative losses in the Balmer emission lines, indicating that the
chromosphere is at least partly heated by the overlying corona.  It is
possible, however, that the somewhat elevated chromospheric luminosity
is the result of an energy input process that took place before the
start of our observations.  This latter possibility is supported by
the apparently decreasing level of peak H$\alpha$ flux during our
observation, and the nearly constant baseline level traced by the
light curve minima.  Indeed, the H$\alpha$ luminosity during the
minima is about half of the X-ray luminosity.

In addition to the persistent emission, we detect a large number of
radio flares with a range of peak fluxes, durations, and degrees of
circular polarization.  The overall short durations of the flares and
large degree of circular polarization are indicative of coherent
emission.  In the context of the electron cyclotron maser mechanism,
the inferred magnetic field is about 3 kG, similar to fields on the
most active early M dwarfs \citep{sl85,jv96}.  Similar flares have
been detected in previous observations of \tvlm, but with a 2 hour
periodicity that is absent in our data.  The 2 hour period was
attributed to compact polar regions in a dipolar field rotating in and
out of our line of sight \citep{hbl+07}.  The durations of the flares
detected here, and their random arrival times and sense of circular
polarization, point instead to a tangled and multi-polar field.  Thus,
the conditions required for coronal coherent radio flares exist on
long timescales, but the change in behavior may signal a shift in the
field configuration on $\lesssim 1$ yr timescales.  The inferred
multi-polar nature of the field is again in good agreement with models
of the $\alpha^2$ dynamo, which suggest that the bulk of the energy is
in the quadrupolar and higher order components.

Unlike in the radio flares, we do find clear H$\alpha$ periodicity
($P\approx 2$ hr), with a sinusoidal light curve that reveals the
presence of a chromospheric hot spot, or an extended bubble, with a
covering fraction of about $50\%$.  It is unclear whether this
emission region is stable over timescales longer than about 1 day, but
the decrease in peak flux between subsequent rotations may point to a
transient nature that may be similar to the one now inferred in the
radio band.  The observed 2-hour period is well matched to the
measured rotation velocity of \tvlm, and indicates a rotation axis
inclination of about 90 deg.  The existence of such an extended
structure provides additional support for a large-scale field that
dominates the quiescent radio and X-ray emission.

The general wisdom in the study of magnetic activity and its impact on
the outer atmosphere is that the input of magnetic energy results in a
series of related events that heat the corona and chromosphere and
result in correlated X-ray, radio, and optical line emission.  This
idea is supported by observations of the Sun, as well as various
samples of early M dwarfs.  The observations presented here show no
clear evidence for any correlation between the various activity bands.
In particular, the quiescent radio emission is over-luminous by nearly
4 orders of magnitude compared to predictions from the radio/X-ray
correlation.  Similarly, the radio flares do not appear to correlate
with the H$\alpha$ variability.  Finally, it is possible that the
X-ray flux incident on the chromosphere is not sufficient to produce
the observed H$\alpha$ luminosity (particularly if we include the
contribution from higher order Balmer lines).

Taking these various observations and inferences into account we
therefore conclude that the observations of \tvlm\ indicate that:

\begin{itemize}

\item Ultracool dwarfs exhibit clear evidence for intense magnetic
activity that is not diminished compared to early M dwarfs.  This
activity is manifested most clearly in the centimeter radio band, but
also in X-rays and Balmer line emission.

\item Both a chromosphere and a corona appear to exist, with
luminosities relative to the bolometric luminosity that are in line
with other objects of similar spectral type, and are significantly
lower (by $\sim 2$ orders of magnitude) compared to early M dwarfs.

\item The dissipation of magnetic energy leads to intense radio
emission, but it does not have a clear effect on chromospheric and
coronal emission, both temporally and in terms of overall luminosity.
The radio/X-ray correlation is violated by about 4 orders of
magnitude.

\item The decoupling of magnetic dissipation (as evidenced by radio 
emission) from chromospheric and coronal emission may be due to
changes in the structure and scale of the magnetic field or the outer
atmosphere.  This is possibly supported by the higher chromospheric
luminosity compared to the coronal luminosity.

\item The presence of a steady large-scale magnetic field, as well as
a multi-polar component, support current models of $\alpha^2$ dynamos
in fully convective stars.

\end{itemize}

Simultaneous multi-wavelength observations of several additional
ultracool dwarfs are in progress.  We expect that with this larger
sample, and with the longer time baselines of our observations
compared to typical studies, we can begin to address in detail the
range of quiescent and variable activity, and the absence or presence
of the correlations that appear to exist in the early M dwarfs.
Ultimately, these observations should reveal the energy and size scale
of the magnetic field, and thus provide a detailed view of the
magnetic generation process in fully convective stars.

\acknowledgements 

We thank the Chandra, Gemini, VLA, and Swift schedulers for their
invaluable help in coordinating these observations.  This work has
made use of the SIMBAD database, operated at CDS, Strasbourg, France.
Based in part on observations obtained at the Gemini Observatory,
which is operated by the Association of Universities for Research in
Astronomy, Inc., under a cooperative agreement with the NSF on behalf
of the Gemini partnership: the National Science Foundation (United
States), the Science and Technology Facilities Council (United
Kingdom), the National Research Council (Canada), CONICYT (Chile), the
Australian Research Council (Australia), CNPq (Brazil) and CONICET
(Argentina).  Data from the UVOT instrument on Swift were used in this
work.  Swift is an international observatory developed and operated in 
the US, UK and Italy, and managed by NASA Goddard Space Flight Center 
with operations center at Penn State University.
Support for this work was provided by the National
Aeronautics and Space Administration through Chandra Award Number
G07-8014A issued by the Chandra X-ray Observatory Center, which is
operated by the Smithsonian Astrophysical Observatory for and on
behalf of the National Aeronautics Space Administration under contract
NAS8-03060.  E.B.~is supported by NASA through Hubble Fellowship grant
HST-01171.01 awarded by the STScI , which is operated by AURA, Inc.,
for NASA under contract NAS 5-26555.


\clearpage

\clearpage

\clearpage

\clearpage

\end{document}